
\magnification=1200
\vskip 0.5cm
\def\dsl{\raise.15ex\hbox{/}\kern-.57em\partial}
\def\Dsl{\,\raise.15ex\hbox{/}\mkern-13.5mu D} 
\def\Asl{\,\raise.15ex\hbox{/}\mkern-13.5mu A} 
\def\Bsl{\,\raise.15ex\hbox{/}\mkern-13.5mu B} 

\font\title=cmr12 at 12pt
\footline={\ifnum\pageno=1\hfill\else\hfill\rm\folio\hfill\fi}
\baselineskip=18pt
\vskip 1.0cm
\centerline{\title MOTION AND TRAJECTORIES OF PARTICLES AROUND}
\centerline{{\title THREE-DIMENSIONAL BLACK HOLES}}
\vskip 2.0cm
\centerline{{\bf C. Farina}$^1$, {\bf J. Gamboa}$^{2,}$
\footnote{$^\dagger$}{\it Also at CECS, Casilla 16443, Santiago 9, Chile.}
and
{\bf Antonio J. Segu\'{\i}-Santonja}$^3$}
\centerline{\it $^1$Instituto de F\'{\i}sica, Universidade
Federal do Rio de Janeiro}
\centerline{\it C.P 68528, BR-21945, R.J., Brazil}
\centerline{\it $^2$Division de Physique Th\'eorique, Institut de Physique
Nucleaire\footnote{$^\ddagger$}{\it Unite de Recherche des Universit\'es
Paris 11 et Paris 6 associ\'ee au CNRS.}}
\centerline{\it F-91406 Cedex, Orsay, France}
\centerline{\it $^3$Departamento de F\'{\i}sica Te\'orica, Universidad
de Zaragoza}
\centerline{\it 50009 Zaragoza, Spain}
\vskip 1.5cm
{\bf Abstract}. The motion of relativistic particles around three dimensional
black holes following the Hamilton-Jacobi formalism is studied. It follows
that the Hamilton-Jacobi equation can be separated and reduced by quadratures
in analogy with the four-\break dimensional case. It is shown that: a)
particles are
trapped by the black hole independently of their energy and angular momentum,
b) matter always falls to the centre of the black hole and cannot
undertake a motion with stables orbits as in four dimensions. For
the extreme values of the angular momentum of the black hole, we were
able to find exact solutions for the equations of motion and
trajectories of a test particle.
\vskip 0.60cm
\leftline{IPNO-TH 93/06}
\leftline{DFTUZ-93/01}
\leftline{February 1993}
\vskip 0.4cm
\leftline{PACS numbers: 04.20.Jb, 97.60.Lf}
\vfill \eject

Since the discovery of three-dimensional gravity
{\bf [1]} many efforts have been performed in order to establish a closer
 analogy between three-dimensional gravity and the four-dimensional case.
Following this perspective, it was shown that in three
dimensions there are classical solutions {\bf [2]} of the Einstein field
equations that keep a narrow relation with the Schwarzschild or Kerr
solutions.
These \lq\lq black hole" (BH) like solutions exhibit a behaviour similar to
their four-dimensional homologous and keep also properties such as horizons or
thermodynamic features {\bf [3]}.
However, in spite of these results, it seems interesting to investigate
to what extent these analogies remain true.
Following this research line, it was found recently {\bf[4]}
other BH-like
solution for a three-dimensional gravity characterized by mass, charge
and angular momentum. Hence, in many aspects, this solution is the natural
 analog of the
four-dimensional Kerr-Newman solution.

The purpose of the present letter is to analyse the motion of relativistic
test particles in the geometry found in {\bf [4]} in order to understand
some issues such as whether there is trapping of particles by this BH or
not, or whether it makes sense to talk about cross section for capturing
particles or not, etc..
This is an interesting point because some of the solutions found in the
literature are solutions with cone-like singularities (exhibiting
deficit angles) and in these cases, there
 is
 no trapping of particles for the same reason by which there is no
trapping of particles by cosmic strings {\bf [5]}.

More precisely we will show
the following issues:
\item{a)} The associated Hamilton-Jacobi equation is separable and trivially
reducible by qua\-dra\-tures.
\item{b)} The two extreme cases, i.e., when the angular momentum of the BH
is zero and maximum respectively, the equations of motion can be
integrated exactly.
\item{c)} The only possible trajectories for test particles, are those
that fall into the singularity.

In order to show these results we start by using the Hamilton-Jacobi formalism,
applied originally by Kaplan and Carter for the Schwarzschild {\bf [6]} and
Kerr
BH {\bf [7]} respectively.

The authors in ref.{\bf [4]} considered the action
$$I={1\over 2\pi}\int d^2x\,\,dt\sqrt{-g}\left[R+2l^{-2}\right] +
B,\eqno(1)$$
where $B$ is a surface term and the radius $l$ is given by
$l={1\over \sqrt{-\Lambda}}$, with $\Lambda$ being the (negative)
cosmological constant. They found that the corresponding Einstein's
equations are solved by the following BH field
$$ ds^2 = -(N^2 - r^2N_\phi)dt^2 + N^{-2}dr^2 + r^2 d\phi^2 +
2r^2N_\phi dt \,d\phi, \eqno(2)$$
where the lapse function $N^2$ and $N_\phi$ are defined as

$$\eqalignno{ &N^2 = -M + {r^2\over l^2} + {J^2\over 4 r^2}, &(3a)  \cr
 &N_\phi = - {J\over 2 {r^2}}. &(3b) \cr}$$
In (3) $M$ and $J$ are two constants of integration that can be understood as
the mass and the angular momentum of the BH respectively.

As it was discussed in {\bf [4]}, the lapse function $N^2$ vanishes for
$$r_\pm = l {\biggl[{M\over 2}\biggl ( 1 \pm
 \sqrt{ 1 - {J^2\over {M^2}{l^2}}}\biggr) \biggr]}^{1\over 2}. \eqno(4)$$
The BH horizon is identified with $r_+$, and it will exist only if
$M$ and $J$ satisfy the relations
$$M>0,\;\;\;\;\; \vert J\vert\leq Ml.\eqno(5)$$
Observe that in the extreme case $\vert J \vert = Ml$ both roots in (4)
coincide.

In order to use the Hamilton-Jacobi equation
$$g^{\mu\nu}{\partial S\over \partial x^\mu}
{\partial S\over \partial x^\nu}+m^2=0,\eqno(6)$$
we see that we need the contravariant components of the metric,
which, after inverting $g_{\mu\nu}$, are given by
$$\eqalignno{ &g^{00} = -N^{-2}, \,\,\,\,\,\,\,\,
\,\,\,\,\,\, g^{11} = N^2, \cr &
 g^{22} = {1\over r^2} (1 - r^2 {N^2_\phi\over N^2}),\,\,\,\,\,\,\,\,\,
g^{02} = g^{20} = {N_\phi\over N^2}. &(7) \cr}$$
\hfill \eject
Substituting (7) in (6), the Hamilton-Jacobi equation for a
relativistic test particle of mass $m$ becomes
$$ -N^{-2}{\biggl({\partial S\over \partial t}\biggr)}^2  +
 N^2 {\biggl({ \partial S\over \partial r}\biggr)}^2 +
{1\over r^2}\left(1 - r^2{N^2_\phi\over N^2}\right)
{\biggl( {\partial S\over \partial \phi}\biggr)}^2
+ 2{N_\phi \over N^2} {\partial S\over \partial t}
{\partial S\over \partial \phi} + m^2 = 0. \eqno(8) $$

The equation (8) can be easily separated with the aid of
 the following ansatz
$$S(r,\phi, t) = -Et + L\phi + S_1(r), \eqno(9)$$
where $E$ and $L$ are the energy and the angular momentum of the particle
respectively.

Substituting (9) into (8) and solving for $S_1$, we get
$$ S_1 = \int {dr\over N} \sqrt{ {\left(E\over N\right)}^2 -
{L^2\over r^2}\left[1 - r^2
{\left({N_\phi\over N}\right)}^2 \right] + 2 {N_\phi E L \over N^2} - m^2 },
 \eqno(10)$$
and a solution of the Hamilton-Jacobi is obtained by quadratures.

The trajectory of the particle can be determined, as usual, by
stating that ${\partial S\over \partial E}$ and
${\partial S\over \partial M}$ are constants respectively. However,
we shall adopt another approach, and we shall obtain the equations of
motion directly from the so called first integral of the geodesic
equation {\bf [8]}
$$P^\mu = m {dX^\mu\over d \tau} = - g^{\mu\nu} {\partial S\over
\partial X^\nu}, \eqno(11)$$
where $\tau$ is the proper-time of the test particle.

Substituting equations (7), (9) and (10) into (11), we obtain, after
straightforward calculations, that
$$\eqalignno{& m{dt\over d\tau} = -{1\over N^2}( E + N_\phi L), &(12a)
\cr &
m{d\phi\over d\tau} = {N_\phi\over N^2} (N_\phi L + E) - {L \over r^2}, &(12b)
\cr &
m^2 {\biggl({dr\over d\tau}\biggr)}^2 =(E+N_\phi L)^2-
\left({LN\over r}\right)^2-m^2N^2. &(12c) \cr}$$

Equations (12a-12c) describe the motion of a relativistic
test particle with mass $m$ in the geometry given by (2). Hence, we
can analyse what kind of motion around the BH can be described by the
test particle.

In order to see whether bounded orbits (which does not mean closed
orbits) are allowed by this geometry or not, the first thing we must
do is to search for turning points. With this goal, we just impose
that the r.h.s. of (12c) must vanish. Hence, the turning points are
given, in principle, by
$$ {\alpha\over r^2} + \beta r^2 +\gamma = 0, \eqno(13)$$
where we defined
$$\eqalignno{ &\alpha = L^2 M - {1\over 4}m^2J^2 - JLE, &(14a) \cr &
\beta = -{m^2\over l^2}, &(14b) \cr &
\gamma = -{L^2\over l^2} + E^2 + m^2 M. &(14c) \cr}$$

Solving (13) we find the roots
$$ \eqalignno{ & R^2_{max} = - {1\over 2\beta} [ \gamma + \Delta], &(15a)
\cr  & R^2_{min} =-{1\over 2\beta}[ \gamma - \Delta], &(15b) \cr}$$
where
$$ \Delta = \sqrt{ \gamma^2 - 4 \alpha \beta}.$$

A necessary condition (but not sufficient) for $R_{min}^2$ to exist
is that $R_{min}^2\geq 0$. Observing that $\beta<0$ by (14b), this
leads to the following condition
$$\alpha \leq 0, \eqno(16)$$
which must always hold.

Once we got expressions for the turning points $R_{min}$ and
$R_{max}$, we could be suggested that both of them would always
exist, or at least that they would exist for some values of the
parameters $M$ and $J$ (which characterizes the BH) and $L$ and $E$
(which characterizes the orbit). This would lead us to the naive
conclusion that bounded orbits are possible for the case at hand.
However, in order to have bounded orbits, we must also be sure that
$R_{min}$ is greater than the BH horizon, and as we shall see, this
will never occur. We shall establish the above result by using two
different approaches.

The first one makes use of the corresponding effective potential,
which can be obtained in the following way: we start in the same way
as if we were interested in obtaining the turning points (let us
denote them by $R_{TP}$), that is, putting the r.h.s. of (12c) equal
to zero. This gives us an algebraic equation like
$F(R_{TP},E,L,M,J)=0$, which, after solving for $R_{TP}$, gives us
the turning points. However, instead of solving for $R_{TP}$, if we
now solve for $E$, and if we consider fixed values for $L,M$ and $J$,
we will get $E=E(R_{TP})$. The effective potential $V_{eff}(r)$ will
coincide precisely with this function, provided $R_{TP}$ is thought
as the variable $r$.

For the problem at hand a direct calculation of
the effective potential yields
$$ V_{eff}(r) = {JL\over 2r^2} +N(r)\sqrt{\left({L\over r}
\right)^2+m^2}, \eqno(17)$$
and a lengthy, but straightforward analysis shows that there are
neither minima nor maxima in the region of physical
interest\footnote{$^1$}{ We have
checked this result numerical and analitically.} (outside the BH horizon).

Hence, independently of its energy and angular momentum, a test
particle always falls into the singularity $r=0$ and
there are no stable orbits. As already mentioned, this result does
not have analog in four dimensions, where some stable orbits are
allowed {\bf [6,9]}.

The same result can be reached by analysing the equations of motion
(12). In fact, we note that although the set of equations (12) is difficult
of to solve, it is possible to extract physical information from them.

Actually, for any value of the BH parameters,  the only
equation that can be exactly integrated is (12c),
$${1\over r^2} = -{2\beta \over { {\gamma + \Delta} \,\,sin \,x}},
\eqno(18)$$
where $x = 2 \sqrt{-\beta}{\tau\over m} $.

{}From this equation we can infer some general consequences
\item{ (a)} When ${\Delta}^2 >0$ the motion of the
particle can be bounded between two circles of radii $R_{max}$ and
$R_{min}$ respectively; if $R_{min}$ is smaller than the horizon such motion
 will exist only
until the particle reachs the gravitational radius where it will be captured
by the BH.

\item{(b)} If ${\Delta}^2 = 0$ the test particle will describe, in
principle, a circular motion and again, it will be
 possible only if its radius is greater than the horizon.

\item{(c)} If ${\Delta}^2 < 0$ the motion will not take place.

The conditions (a)-(b) imply relations between
the parameters of the particle $(m,L,E)$ and the BH ones $(M,J)$.
For a circular motion this relation is
$$J = {{(El - L)}^2 \over m^2l} + Ml, \eqno(19)$$

However, $Ml$ is the maximum angular momentum of the BH which is
physically allowed and, as a consequence, a circular orbit is possible
only if both relations $J=Ml$ and $El=L$ are simultaneously satisfied.

On the other hand, the radius $R_0$ of this circular orbit is given
by (it suffices to substitute $\Delta=0$ into (15))
$${R_0}^2 = {-\gamma \over 2 \beta}\vert_{J=Ml={ML\over E}} = {Ml^2 \over 2} =
{R_g}^2, \eqno(20)$$
which is just the horizon (4) in the extreme case $J=Ml$. Thus, we conclude
that, although equation (18) gives an oscillating solution, the
presence of the horizon implies that a circular motion can never
occur, which means that there is no minimum of $V_{eff}(r)$ in the
physical region.

Therefore, we have seen by using two different points of view that
the three-di\-men\-sio\-nal BH just studied are objects that, although
trap matter, never allow stable orbits for test particles moving
around them.

Now let us discuss some particular cases where we were able to obtain
exact solutions of the equations (12).
\vskip 0.15cm
${\it a)\,\, \underline {Case \,\,J=0}}$

This situation corresponds to have an anti- de Sitter metric provided
we perform the transformation $t \to {t\over \sqrt{-M}}$ in the metric (2).
It is interesting to note that for this case (16) imply $M<0$ and, as a
consequence, the limit $J \to 0$ corresponds precisely the solution
discussed in {\bf [1]}.
The solutions of the corresponding equations of motion can be written
as
$$\eqalignno{ & {1\over r^2} = -{2\beta \over { \gamma + \Delta^{'} sin\,\,
x}},
 &(21a)
\cr &
 {\Delta}^{'} + \gamma\,\, tan {x\over 2} = -2\sqrt{\rho\,\,\beta}\,\,
tan ({\sqrt{-\rho}\over L}\phi ), &(21b) \cr}$$
while the trajectory is given by
$${\gamma + 2 \beta\, r^2\over \Delta^{'}} =
2\gamma\,{\Delta^{'} + 2\sqrt{\rho \, \beta} \,\,
\tan\,\, ({\sqrt{-\rho}\over L} \phi)\over
\gamma^2 + {\left[ \Delta^{'} + 2\sqrt{\rho\, \beta}\,\,
\tan \,\,({\sqrt{-\rho}\over L}\phi)\right]}^2}, \eqno(22)$$
with $\Delta^{'} = \sqrt{\gamma^2 - 4 \rho \beta}$ and $\rho = ML^2$.

Some geometrical properties of this solution are discussed, for instance,
in {\bf [10-11]}.
\vskip 0.15cm
${\it b)\,\, \underline{ Case\,\, J= Ml}}$

This extreme case is interesting because the horizon still
remains and the equations (12) are notably simplified.

The solutions of the equations of motion are given by
$$\eqalignno{  &{1\over r^2} = -{2 \beta \over {\gamma + \Delta \,\, sin\, x}}
\cr &
{1\over \sqrt{-\beta}}\phi = {1\over {a^2 - b^2}}
\biggl[ {{(Ab - Ba)}\,\, \cos\,\,x \over {a + b\,\, \sin\,\, x}} +
{(Aa - Bb)}{\cal J}
\biggr], &(23)
\cr & {1\over \sqrt{-\beta}}\,\, t = {1\over {a^2 - b^2}}
\biggl[ {{(Ab -aB)}\,\,\cos\,\, x\over {a + b\,\, \sin\,\, x}}  +
{(A^{'} a - B^{'}b)} {\cal J} \biggr], \cr}$$
where the coefficients are defined as
$$\eqalignno{ & A = 2\beta L M l^2 - Ml^3 E\beta + L \gamma,\,\,\,\,\,
B =  L \Delta,
\cr &
A^{'} = l^2 E \gamma + Ml^3 \beta,\,\,\,\,\,
B^{'} = l^2 E \Delta,
\cr &
a = \gamma + \beta Ml^2, \,\,\,\,\,
b = \Delta, \cr}$$
and
$${\cal J} = \cases{{2\over \sqrt{a^2 - b^2}} {\tan}^{-1} \left[
{{a \tan{x\over 2} + b}\over {\sqrt{a^2 - b^2}}} \right] &if
  $a^{2} > b^{2}$
\cr
{1\over {\sqrt{b^{2} - a^{2}}}} \ln \left[
{{a\,\tan{x\over 2} + b - {\sqrt{b^{2} - a^{2}}}}\over {a\, \tan{x\over 2} +
b + \sqrt{b^{2} - a^{2}}}} \right] &if $a^{2} < b^{2}$ .\cr}$$

The equation of the trajectory is tedious to write, so we avoid this
calculation since we shall not need it for future applications.

For the case of a charged BH, our results can be extended
straightforwardly. In fact, as it was discussed in {\bf [4]} the
electromagnetic coupling to
the metric (2) corresponds to perform the change
$$N^2 \to N^2 + {Q^2\over 2} \ln \left({r\over r_0} \right), \eqno(24)$$
being $Q$ the electric charge of the BH and $r_0$, a constant.

The new effective potential $V^Q_{eff}$ for this case has the same structure
of that given by (17), provided the replacement (24) is performed. Again,
$V^Q_{eff}$ does not have neither maxima nor minima and the same previous
conclusions for the uncharged BH can be reached. However, for the
charged BH, the equations of motion become very
complicated to be solved analytically. This last example shows that
the effective potential approach may be sometimes more convenient
than trying to solve directly the corresponding equations of motion.

We would like to thank M. Asorey by discussions a critical
reading of the manuscript.
C.F. would like to thank to F.M. de Almeida, R. Donangelo and
L.S. de Paula for some help with internet problems and also to the
Department of Theoretical Physics of Zaragoza where part of this work
was done. J.G. thanks J. Zanelli by discussions and correspondence
concerning this work. This research was partially supported by
CNPq-Brazil (C.F) by CNRS and FONDECYT grant
0867/91 (J.G) and by DGICYT (Spain) grant PB90-0916 (A.J. S.-S).
\vskip 0.5cm
\centerline{\bf References}

\item{{\bf [1]}} S. Deser, R. Jackiw and G. 't Hooft, {\it Ann. of Phys.}
(N.Y.) {\bf 152}, 220(1984).
\item{{\bf [2]}} G. Cl\'ement, {\it Int. J. Theor. Phys.} {\bf 24}, 267(1985).
\item{{\bf [3]}} B. Reznik, {\it Phys. Rev.} {\bf D45}, 2151(1992).
\item{{\bf [4]}} M. Ba\~nados, C. Teitelboim and J. Zanelli, {\it
Phys. Rev. Lett.} {\bf 69}, 1819(1992).
\item{{\bf [5]}} J. Gamboa and A. Segu\'{\i}-Santonja, {\it Class. and Quant.
Grav.} {\bf 9}, L111(1992).
\item{{\bf [6]}} S.A. Kaplan, {\it Zh. Eksp. Fiz.} {\bf 19}, 951(1949).
\item{{\bf [7]}} B. Carter, {\it Phys. Rev.} {\bf 174}, 1559(1968).
\item{{\bf [8]}} L.D. Landau and E.M. Lifshitz, {\it Classical Theory of
Fields}, Pergamon Press (1975).
\item{{\bf [9]}} D.C. Wilkins, {\it Phys. Rev.} {\bf D25}, 814(1972).
\item{{\bf [10]}} J.D. Brown and M. Henneaux, {\it Comm. Math. Phys.}
{\bf 104}, 207(1986).
\item{{\bf [11]}} J.D. Brown, {\it Lower Dimensional Gravity},
World Scientific (1989).
\end